\documentclass[11pt]{article}

\usepackage{amsmath}
\usepackage{graphicx}
\usepackage{indentfirst}
\usepackage{amssymb}
\usepackage{cite}
\usepackage{color}
\usepackage{subfigure}

\begin{document}

\title{\bf{Charged Black Hole in Gravity's Rainbow:\\Violation of Weak Cosmic Censorship}}

\date{}
\maketitle

\begin{center}
\author{Yongwan Gim}$^{a}$\footnote{yongwan89@sogang.ac.kr} and
\author{Bogeun Gwak}$^{b}$\footnote{rasenis@sejong.ac.kr}\\

\vskip 0.25in
$^{a}$\it{Department of Physics and Research Institute for Basic Science, Sogang University, Seoul 04107, \\ Republic of Korea}\\
$^{b}$\it{Department of Physics and Astronomy, Sejong University, Seoul 05006, Republic of Korea}

\end{center}
\vskip 0.6in
{\abstract
{
We investigate the validity of weak cosmic censorship conjecture for electrically charged black holes in the presence of gravity's rainbow under charged particle absorption. The rainbow effect is shown to play an important role when the black hole is modified by a particle carrying energy and electric charge. Remarkably, we prove that the rainbow-charged black hole can be overspun beyond the extremal condition under charged particle absorption. Further, it is demonstrated that the second law of thermodynamics and cosmic censorship conjecture are violated owing to the rainbow effect.}
}

\thispagestyle{empty}
\newpage
\setcounter{page}{1}

\section{Introduction}\label{sec1}

Black holes are one of the interesting topics related to various phenomena, such as, gravitational waves and gamma-ray burst. Further, the structure of a black hole is very different compared with other stellar objects. This is because the event horizon covers the inside of black holes. No particle, not even light, can escape once it is inside a black hole after passing through the horizon. When a particle enters a black hole, its irreducible mass increases as an extensive property \cite{Christodoulou:1970wf,Christodoulou:1972kt}, which is an energy distributed on the surface of the horizon\cite{Smarr:1972kt}. On the contrary, the black hole has a reducible energy in the form of rotational and electric energies. Reducible energy can decrease by a specific process, such as, the Penrose process\cite{Bardeen:1970zz,Penrose:1971uk}. The extensive behavior of an irreducible mass is understood in terms of black hole thermodynamics, and the square of an irreducible mass is proportional to the Bekenstein--Hawking entropy of a black hole\cite{Bekenstein:1973ur,Bekenstein:1974ax}. Further, as a conjugate variable of the entropy, the Hawking temperature is obtained from an emission through a quantum effect on the horizon\cite{Hawking:1974sw,Hawking:1976de}. Hence, the black hole can be studied as a thermal system following the laws of thermodynamics.

There is an interesting conjecture related to the internal structure of a black hole. A singularity is located at the center of the black hole spacetime that is covered by the horizon; thus, it cannot be seen by an external observer. However, in the absence of the horizon, the naked singularity is exposed, and causality breaks down. To prevent this, the weak cosmic censorship conjecture ensures that the horizon always stably covers the singularity located inside a black hole\cite{Penrose:1964wq,Penrose:1969pc}. However, there is no general proof on the validity of the conjecture for black holes; therefore, we need to test the conjecture for each type of black hole. The first investigation considered a Kerr black hole, which cannot be overspun beyond the extremal condition by adding a particle \cite{Wald:1974ge}. Further, the self-force effect of a particle is shown to be important in ensuring the stability of the horizon in the Kerr black hole\cite{Barausse:2011vx,Colleoni:2015ena}. Moreover, a charged black hole was also considered in the weak cosmic censorship conjecture, because a backreaction was examined in \cite{Hubeny:1998ga,Isoyama:2011ea}. Several tests on various black holes included in modified gravity theories are still being performed\cite{Semiz:2005gs,Gao:2012ca,Zhang:2013tba,Rocha:2014jma,McInnes:2015vga,Cardoso:2015xtj,Gwak:2016gwj,Horowitz:2016ezu,Revelar:2017sem,Duztas:2017lxk,Song:2017mdx,Gwak:2018akg,Yu:2018eqq,Liang:2018wzd,Zeng:2019jrh,Han:2019kjr,Zeng:2019jta,Chen:2019pdj,Han:2019lfs,Zeng:2019aao}. In addition, we found that the validity of the cosmic censorship conjecture is closely related to the first and second law of thermodynamics for a given black hole system\cite{Gwak:2015fsa}. Hence, if the first law is assumed to have the thermodynamic volume and pressure term in a charged anti-de Sitter black hole, the second law and weak cosmic censorship conjecture are significantly affected\cite{Gwak:2017kkt,Gwak:2019asi}.

On the other hand, considerable attention has been paid to modified dispersion relations (MDRs). Many efforts have been devoted to studying the various aspects of MDRs\cite{ Aloisio:2005qt,  Girelli:2006fw, Garattini:2011kp, Majhi:2013koa, Kiyota:2015dla,Rosati:2015pga, Barcaroli:2015xda, Barcaroli:2016yrl}. In this context, it was suggested that the Minkowski spacetime should be deformed by the energy of a particle laid on the spacetime. Gravity's rainbow is a generalization of this deformation to present MDRs to curved spacetimes, such as, black holes and cosmology. Hence, a geometry is distorted by the energy of the test particle moving in it. The concepts of Schwarzschild black hole and Friedmann--Robertson--Walker cosmology were first modified in terms of gravity's rainbow\cite{Magueijo:2002xx}. Gravity's rainbow has also been studied to investigate various aspects of black holes\cite{Galan:2006by, Garattini:2012ec, Ali:2014xqa, Gim:2014ira, Ali:2014qra, Gim:2015zra, Gim:2015yxa, Hendi:2016wwj, Hendi:2016hbe, EslamPanah:2017ugi, Gim:2017rmn, Hendi:2018sbe} and cosmology\cite{Ling:2006az, Ling:2008sy, Barrow:2013gia, Awad:2013nxa, Chang:2014tca, Garattini:2014rwa, Khodadi:2015tsa, Hendi:2017vgo}. 
Note that the deformation owing to the rainbow effect leads to Lorentz violations; however, it explains phenomena such as threshold anomalies in TeV photons and ultra-high cosmic rays \cite{AmelinoCamelia:1996pj, AmelinoCamelia:1997gz, AmelinoCamelia:1997jx, Colladay:1998fq, Coleman:1998ti, AmelinoCamelia:1999wk, AmelinoCamelia:2000zs, Jacobson:2001tu, Jacobson:2003bn, AmelinoCamelia:2008qg}. However, the weak cosmic censorship conjecture has not been thoroughly studied for rainbow black holes.

In this work, we investigate the validity of the weak cosmic censorship conjecture for electrically charged black holes in the presence of gravity's rainbow under charged particle absorption. Particularly, the weak cosmic censorship conjecture is not well studied in the presence of gravity's rainbow; hence, this study is the first demonstration of the weak cosmic censorship conjecture in the presence of gravity's rainbow. Here, we set a specific pair of rainbow functions; the results agree with those of the quantum-spacetime-phenomenology perspective \cite{AmelinoCamelia:1996pj, AmelinoCamelia:1997gz} and loop-quantum-gravity approach\cite{Gambini:1998it, Alfaro:2001rb, Sahlmann:2002qk, Smolin:2002sz, Smolin:2005cz}. Because the metric of the black hole is deformed owing to the rainbow effect, we can elucidate the influence of the deformation on the weak cosmic censorship conjecture and thermodynamics in the black hole. Particularly, because the charged black hole without the rainbow effect is already studied in the weak cosmic censorship conjecture, our investigation is the first analysis to show the differences between with and without the rainbow effect. Furthermore, since the violation of the weak cosmic censorship conjecture was shown in the five-dimensional black ring with an asymptotically flat spacetime\cite{Figueras:2015hkb}, our analysis can demonstrate that the rainbow effect can play an important role in the violation. The conjecture is investigated for a charged particle entering the black hole. Interestingly, the initial state is the rainbow charged black hole deformed by gravity's rainbow originating from the charged particle, and the final state is the charged black hole without the rainbow effect because there is no particle in the final state. Non-extremal black holes and extremal black hole are considered in their initial states. Further, we determine whether the extremal black hole can spin beyond the extremal condition under charged particle absorption in the presence of gravity's rainbow.

This paper is organized as follows. In section\,\ref{sec2}, we review charged black holes in Einstein's gravity coupled with Maxwell field. Then, gravity's rainbow and rainbow charged black holes are introduced. In section\,\ref{sec3}, we obtain a solution to Hamilton--Jacobi equations for a charged particle in the rainbow charged black hole. Then, changes in the outer horizon and Bekenstein--Hawking entropy are investigated in cases of non-extremal black holes. In section\,\ref{sec4}, we investigate the weak cosmic censorship conjecture for extremal black holes. In section\,\ref{sec5}, we briefly summarize our results.

\section{Charged Black Holes in Gravity's Rainbow}\label{sec2}

In this work, we consider the charged black hole spacetime modified by the energy of a charged particle. If the rainbow effects of the particle are not considered, charged black holes are asymptotically flat solutions to Einstein's gravity coupled with Maxwell field $F_{\mu\nu}$. The action is expressed as
\begin{align}\label{eq:S_RN}
S = \frac{1}{16\pi } \int  d^4 x \sqrt{-g} ( R - F_{\mu\nu}F^{\mu\nu} ),
\end{align}
where $R$ is the curvature. The field equations are
\begin{align}\label{eq:EOM_RN}
\nabla F^{\mu\nu}=0~, 
\quad R_{\mu\nu}-\frac{1}{2}g_{\mu\nu}R = 2 \left(F_{\mu\rho}F^{\rho}_\nu -\frac{1}{4}g_{\mu\nu}F_{ \lambda\rho}F^{\lambda\rho}\right). 
\end{align}
The charged black hole is a spherical symmetric solution to Eq.\,(\ref{eq:EOM_RN}). The metric of the charged black hole is of mass $M$ and electric charge $Q$ as
\begin{align}\label{eq:E_metric}
ds^2=-H(r)d t^2 +\frac{1}{H(r)}dr^2 + r^2(d\theta^2 +\sin^2\theta d\phi^2), \quad H(r)=1- \frac{2 M}{r} + \frac{Q^2}{r^2}, \quad A=- \frac{Q}{r} dt,
\end{align}
where $A$ is the electric potential. The inner and outer horizons of the charged black hole are expressed as
\begin{align}
r_\text{i} = M - \sqrt{M^2 - Q^2},\quad r_\text{h} = M + \sqrt{M^2 - Q^2}.
\end{align}
In addition, the extremal condition is $M=Q$, where the electric charge is maximized for a given mass. The singularity of the spacetime is estimated using the Kretschmann scalar.
\begin{align}
R^{\mu\nu\sigma\rho} R_{\mu\nu\sigma\rho}=\frac{56Q^4}{r^8}-\frac{96 M Q^2}{r^7}+\frac{48 M^2}{r^6},\nonumber
\end{align}
which diverges at the origin, $r=0$. Hence, we ensure that the singularity is located there. We mainly consider thermodynamics on the outer horizon; hence, all thermodynamic properties are defined on it. Then, the surface area of the outer horizon $A_\text{h}$ and Bekenstein--Hawking entropy $S_{\rm RN}$ are related as
\begin{equation}\label{eq:T_RN}
S_{\rm RN} = \frac{A_{\rm h}}{4} = \pi r_\text{h}^2.
\end{equation}
The Hawking temperature $T_\text{RN}$ and electric potential $\Phi_\text{RN}$ of the charged black hole are given by
\begin{equation}\label{eq:phi1st_RN}
T_{\rm RN} = \frac{r_\text{h}-M}{2\pi r_\text{h}^2}, \quad \Phi_{\rm RN} = \frac{Q}{ r_\text{h}}.
\end{equation}
Then, the thermodynamic variables are related by the first law of thermodynamics as
\begin{equation}\label{eq:1st_RN}
dM = T_{\rm RN} dS_{\rm RN} + \Phi_{\rm RN } dQ,
\end{equation}
which determines the change in the charged black hole under the infinitesimal variation.

Then, we consider a charged particle moving in the charged black hole spacetime. Because the energy and momentum of the particle can affect the spacetime structure in consideration of MDRs, we should consider the effects of the MDR to obtain more precise results. Gravity's rainbow considers the MDR of the particle. Here, the MDR is given by\cite{AmelinoCamelia:1997gz,Magueijo:2002xx}
\begin{align}\label{eq:MDR}
f(E)^2 E^2-g(E)^2 p^2 = m^2,
\end{align}
where $E,~p,~m$ represent the energy, momentum, and mass of a test particle, respectively. This is the dispersion relation for the particle in the asymptotically flat region. Hence, this shows the modification owing to the rainbow effect in the flat spacetime.  Furthermore, there are two independent variables: $E$ and $p$, because of the dispersion relation. The effect of gravity's rainbow is imposed by functions $f(E)$ and $g(E)$, which denote rainbow functions. The rainbow functions should be reduced to $\lim_{E\rightarrow 0} f =1$ and $\lim_{E\rightarrow 0} g =1$, because the effect of the MDR is consistent with that of the ordinary dispersion relation in the low energy limit. Note that the MDR \eqref{eq:MDR} can be rewritten as the ordinary dispersion relation of $\tilde{E}^2-\tilde{p}^2=m^2$ in terms of rescaled energy and momentum of the particle considering the rainbow effect. According to gravity's rainbow, rescaled energy and momentum are related and transformed from $\tilde{E}=f(E)E$ and $\tilde{p}=g(E)p$.

The MDR in Eq.\,(\ref{eq:MDR}) is about a particle moving in Minkowski spacetime. Hence, we need to generalize the MDR to that of the curved spacetime of a black hole as done in \cite{Magueijo:2002xx}. Because the MDR of the black hole should be coincident with Eq.\,(\ref{eq:MDR}) in its asymptotically flat limit, the metric of the black hole is also modified to include the rainbow function. Compatible with Eq.\,(\ref{eq:MDR}), the metric of a black hole with gravity's rainbow is obtained under the transformation based on \cite{Magueijo:2002xx}:
\begin{align}\label{eq:tandr}
\tilde{t}(E) =\frac{t}{f(E)}, \quad \tilde{r}(E)= \frac{r}{g(E)}, \quad d\tilde{t}(E) =\frac{dt}{f(E)}, \quad d\tilde{r}(E)= \frac{dr}{g(E)}, \quad  \tilde{G}(E) = \frac{G}{g(E)}.
\end{align}
Under the transformation in Eq.\,(\ref{eq:tandr}), the metric of the charged black hole in Eq.\,(\ref{eq:E_metric}) becomes
\begin{align}\label{eq:metric_final}
ds^2=-\frac{F(r)}{f(E)^2}dt^2 +\frac{1}{F(r) g(E)^2}dr^2 + \frac{r^2}{g(E)^2}(d\theta^2 +\sin^2\theta d\phi^2),
\end{align}
where we omit the tilde signs to avoid confusion. The function $F(r)$ is $H(r)$ of Eq.\,(\ref{eq:E_metric}) modified by the rainbow effect. The function $F(r)$ and electric potential are expressed as
\begin{align}\label{eq:F(r)andA}
F(r)=1- \frac{2GM}{r} + \frac{g(E)GQ^2}{r^2}, \quad A=-\frac{g(E)}{ f(E)} \frac{Q}{r} dt.
\end{align}
Because the rainbow charged black hole in Eq.\,(\ref{eq:metric_final}) includes energy dependence on its metric, various thermodynamic properties are imposed on the rainbow effect. The location of the inner and outer horizons $r_{\rm I}$ and $r_{\rm H}$ are given by
\begin{align}\label{eq:horizon07}
r_{\rm I}= GM-\sqrt{G^2 M^2 - G Q^2 g(E)},\quad r_{\rm H}= GM+\sqrt{G^2 M^2 - G Q^2 g(E)}.
\end{align}
Then, in the rainbow charged black hole, the extremal condition and the horizon become
\begin{align}
M=\sqrt{\frac{g(E)}{G}}Q,\quad r_\text{H}=G M.
\end{align}
Owing to the rainbow effect, the Kretschmann scalar is deformed to
\begin{align}
R^{\mu\nu\sigma\rho} R_{\mu\nu\sigma\rho}=\frac{56 g(E)^6 Q^4}{r^8}-\frac{96 g(E)^5 M Q^2}{r^7}+\frac{48 g(E)^4 M^2}{r^6},\nonumber
\end{align}
which still diverges at the origin of the spacetime. Hence, despite the rainbow effect, the singularity exists at the origin. The Bekenstein--Hawking entropy of the rainbow charged black hole is calculated as\cite{Hendi:2018sbe}
\begin{align}\label{eq:TH}
S_{\rm H} = \frac{\pi r_{\rm H}^2}{g(E)^2 G}. 
\end{align}
We set $G=1$ to avoid confusion as follows.

The rainbow effect will be realized when we consider the forms of the rainbow functions. Various forms of rainbow functions are allowed in gravity's rainbow. Here, we consider the black hole spacetime modified by the rainbow effect originated from the charged particle. Hence, from a quantum-spacetime phenomenology perspective\cite{AmelinoCamelia:1996pj, AmelinoCamelia:1997gz}, the MDR is appropriate in our case. The form of the MDR is given as \cite{AmelinoCamelia:1996pj, AmelinoCamelia:1997gz, AmelinoCamelia:2008qg}
\begin{align}\label{eq:MDR2}
m^2 \approx E^2 - p^2 + \eta  \frac{E^n}{E_{\rm p}^{n}} p^2,
\end{align}
where $E_{\rm P}$ denotes the Planck energy, $\eta$ is a positive free parameter, and $n$ is a positive integer. Further, the MDR agrees with results from the loop-quantum-gravity approach \cite{Gambini:1998it, Alfaro:2001rb, Sahlmann:2002qk, Smolin:2002sz, Smolin:2005cz}.
By comparison with Eqs.~\eqref{eq:MDR} and \eqref{eq:MDR2}, the rainbow functions are expressed as
\begin{align}\label{eq:fandg}
f(E) =1, \qquad g(E) = \sqrt{1-\eta \frac{E^n}{E_{\rm P}^n}},
\end{align}
where $n =1$ will be chosen for analytic calculations.

\section{Charged Particle Absorption in Rainbow Charged Black Hole}\label{sec3}

We consider variations of the rainbow charged black hole caused by charged particle absorption. When a particle enters a black hole, its conserved quantities will be transferred to those of the black hole. The mass and electric charge of the black hole are assumed to vary as much as the energy and electric charge carried by the particle at the outer horizon. The conserved quantities of the particle will be obtained in terms of a dispersion relation obtained by solving the equations of motion of the particle. To derive the dispersion relation of the particle in the rainbow charged black hole, we will apply the Hamilton--Jacobi method and separate variable. The Hamiltonian of a particle having an electric charge $q$ in the electric potential $A_\mu$ is expressed as
\begin{align}\label{eq:H}
{\cal H} = \frac{1}{2}g^{\mu\nu}(p_\mu-q A_\mu)(p_\nu - qA_\nu),
\end{align}
where the four momentum $p_\mu$ of the particle is defined as
\begin{align}\label{eq:pmu}
p_\mu = \partial_\mu {\cal I}.
\end{align}
The Hamilton--Jacobi action ${\cal I}$ of the charged particle is given by
\begin{align}\label{eq:I}
{\cal I} = \frac{1}{2}m^2 \lambda - E t  +{\cal I}_r(r) + {\cal I}_\theta(\theta) +L \phi.
\end{align}
where $\lambda$ is the affine parameter. According to translation symmetries to $t$ and $\phi$ coordinates in the metric Eq.\,(\ref{eq:metric_final}), $E$ and $L$ are defined as the energy and angular momentum, respectively. However, there are no translation symmetries to radial and $\theta$-directional coordinates; therefore, the specific forms for these coordinates cannot be clarified. Instead, we denote them as ${\cal I}_r(r)$ and ${\cal I}_\theta(\theta)$. These parameters can be obtained in terms of the conserved quantities and a separate variable. Then, from Eqs.~\eqref{eq:H} and \eqref{eq:I}, the Hamiltonian equation becomes
\begin{align}\label{eq:HJeq}
- 2 \frac{\partial {\cal I}}{\partial \lambda} =& -\frac{f(E)^2}{F(r)}\left(-E+\frac{g(E)}{f(E)}\frac{q Q}{r}\right)^2 + g(E)^2 F(r) (\partial_r {\cal I}_r )^2 + \frac{g(E)^2}{r^2}(\partial_\theta {\cal I}_\theta)^2 +\frac{g(E)^2}{r^2 \sin^2\theta}L^2\\
=&-m^2.\nonumber
\end{align}
Using a separate variable $K$, the Hamiltonian equation in Eq.\,(\ref{eq:HJeq}) is divided into radial and $\theta$-directional equations\cite{Carter}.
\begin{align}
K &= -\frac{m^2 r^2}{g(E)^2} +\frac{r^2 f(E)^2}{g(E)^2 F(r)  }\left(-E+\frac{g(E)}{f(E)}\frac{q Q}{r}\right)^2 - r^2 F(r) (\partial_r {\cal I}_r )^2,
\quad K =  (\partial_\theta {\cal I}_\theta)^2 +\frac{1}{ \sin^2\theta}L^2.
\end{align}
The Hamilton--Jacobi action in Eq.\,(\ref{eq:I}) becomes
\begin{align}\label{eq:I02}
\mathcal{I}=\frac{1}{2}m^2 \lambda -Et +\int dr \sqrt{R} +\int d\theta \sqrt{\Theta} +L\phi,
\end{align}
where
\begin{align}
\mathcal{I}_r &= \int dr \sqrt{R}, \quad R=\frac{1}{r^2 F(r)}\left(-K-\frac{m^2 r^2}{g(E)^2} +\frac{r^2 f(E)^2}{g(E)^2 F(r)  }\left(-E+\frac{g(E)}{f(E)}\frac{q Q}{r}\right)^2\right),\\
\mathcal{I}_\theta &= \int d\theta \sqrt{\Theta}, \quad \Theta=K-\frac{1}{\sin^2\theta}L^2.\nonumber
\end{align}
By solving Eq.\,(\ref{eq:I02}), the radial and $\theta$-directional angular momentum of the charged particles for a given location are obtained as
\begin{align}
p^r  &= g(E)^2  \sqrt{ -\frac{F(r) }{r^2 }K +\frac{F(r) }{ g(E)^2}m^2 +\frac{ f(E)^2}{g(E)^2 }\left(-E+\frac{g(E)}{f(E)}\frac{q Q}{r}\right)^2}, \label{eq:pr} \\
p^\theta &= \frac{g(E)^2}{r^2} \sqrt{K -\frac{1}{ \sin^2\theta}L^2}.\nonumber
\end{align}
By removing $\mathcal{K}$ in Eq.~\eqref{eq:pr}, the momenta and conserved quantities of the charged particle are related to the dispersion relation as 
\begin{align}\label{eq:DR_given_r}
f(E)^2\left(-E + \frac{ q Q g(E)}{r f(E)}\right)^2 = \frac{(p^r)^2}{g(E)^2} + F(r)\left(-m^2+\frac{r^2}{g(E)^2}(p^\theta)^2+\frac{g(E)^2}{r^2\sin^2\theta}L^2\right).
\end{align}
A charged particle passing through the outer horizon $r_{\rm H}$ is assumed to be completely absorbed by the black hole, because the particle is indistinguishable from the black hole as seen by an observer outside the horizon. Hence, at the outer horizon, the energy and electric charge of the particle contribute to the black hole. In the limit to the outer horizon, the dispersion relation \eqref{eq:DR_given_r} becomes 
\begin{align}\label{eq:DR1}
E- \frac{g(E)}{f(E)}\frac{qQ}{r_{\rm H}} = \frac{1}{f(E)g(E)}|p^r|,
\end{align}
where we choose the positive sign in front of the kinetic term $|p^r|$ in Eq.\,(\ref{eq:DR1}). This choice positively relates the energy of the particle to its kinetic energy without the electric potential of the $Q$ term in the positive flow of time \cite{Christodoulou:1970wf, Christodoulou:1972kt}. The dispersion relation in Eq.\,(\ref{eq:DR1}) is for the particle near the outer horizon of the black hole with the rainbow effect. The total energy of the particle can be negative with the contribution of the electric potential, when the electric attraction acts on the particle.

We now investigate the variation of the rainbow charged black hole under charged particle absorption. Here, there are conserved quantities of the black hole and the particle. Particularly, the mass and electric charge of the black hole are assumed to change owing to the conserved quantities carried by the charged particle. Furthermore, the energy and electric charge of the particle are also conserved quantities that cannot disappear in the spacetime. Hence, these quantities are assumed to be carried into those of the black hole and still conserved under absorption\cite{Christodoulou:1970wf,Christodoulou:1972kt}. Then, the energy and electric charge of the charged particle change the mass and electric charge of the black hole. We assume that
\begin{align}
dM=E,\quad dQ=q.
\end{align} 
The energy and electric charge are related by Eq.\,(\ref{eq:DR1}). Hence, the changes in the mass and electric charge of the black hole are also constrained by the relation
\begin{align}\label{eq:DR2}
f(dM) dM = \frac{g(dM)Q}{r_{\rm H}}dQ + \frac{1}{g(dM)}|p^r|.
\end{align}
The variables originated from the charged particle such as $dM$, $dQ$, and $p^r$ are infinitesimally small compared with the mass $M$ and charge $Q$ of the black hole; therefore, $dM, dQ, |p^r| \ll M, Q$. Then, we can expand Eq.~\eqref{eq:DR2} with respect to the variables $dM$, charge $dQ$, and momentum $|p^r|$ by substituting the horizon $r_{\rm H}$ of the rainbow black hole in Eq.\,(\ref{eq:horizon07}). Since the second orders of these variables are sufficiently small and can be neglected, the dispersion relation in Eq.\,(\ref{eq:DR2}) can be obtained in terms of the first orders of $dM$, $dQ$, and $p^r$. Solving the dispersion relation about the change in the mass, we obtain
\begin{align}\label{eq:dispersion09}
dM=|p^r|+\frac{Q}{M+\sqrt{M^2-Q^2}}dQ,
\end{align}
which has the same form as the ordinary dispersion relation of the charged black hole without the rainbow effects: $f(dM) =1$ and $g(dM)=1$. However, qualitatively, Eq.\,(\ref{eq:dispersion09}) is different from the ordinary one, because, here, $M=Q$ is not the extremal condition. The extremal condition of Eq.\,(\ref{eq:dispersion09}) is at $M=g(dM)Q$. The variables are defined at the rainbow charged black hole.

\begin{figure}[h]
  \begin{center}
\subfigure[{~$Q-\frac{q}{|p^r|}$ diagram for $\eta = 0.02$}]{
  \includegraphics[width=0.321\textwidth]{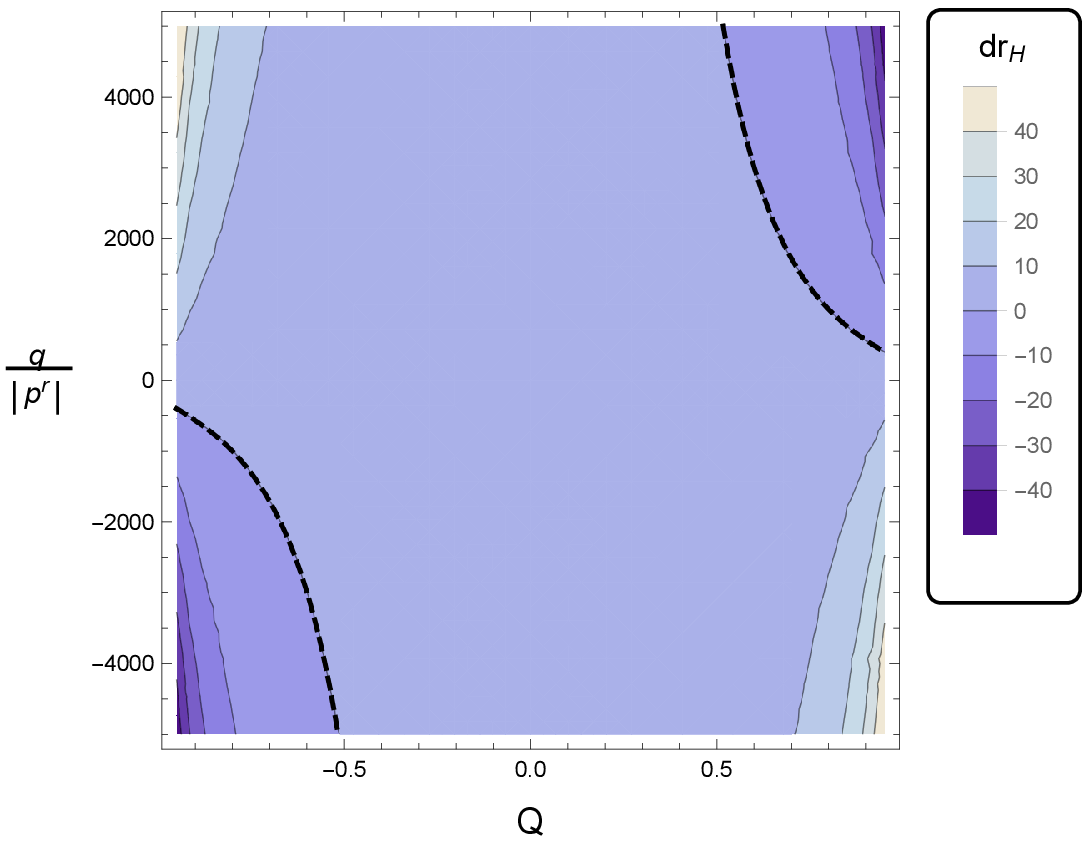}\label{fig:1kkaa1}}
\subfigure[{~$Q-\frac{q}{|p^r|}$ diagram for $\eta = 0.5$}]{
 \includegraphics[width=0.321\textwidth]{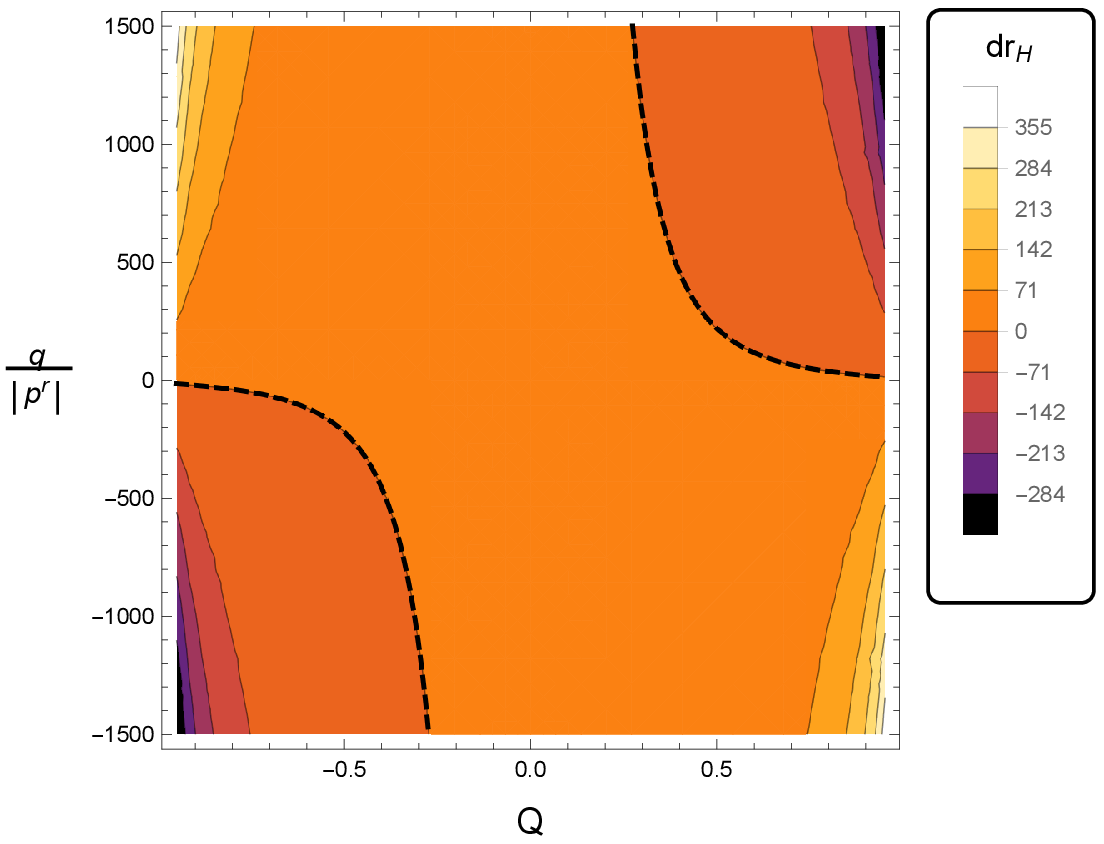}\label{fig:1kka2}}
\subfigure[{~$Q-\frac{q}{|p^r|}$ diagram for $\eta = 0.9$}]{
 \includegraphics[width=0.321\textwidth]{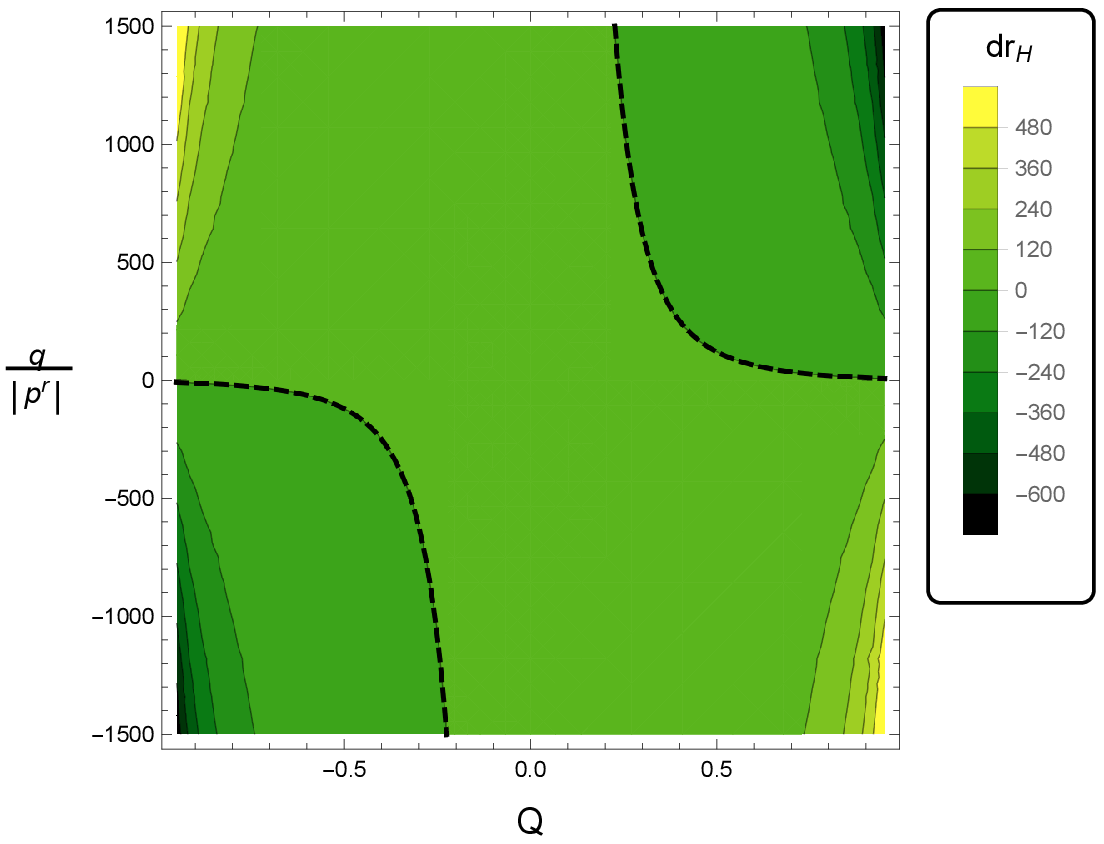}\label{fig:1kkaa3}}
  \end{center}
  \caption{Changes in the outer horizon $dr_\text{H}$ in $Q-\frac{q}{|p^r|}$ diagrams of $M=1$ for a given $\eta$.}\label{fig:fig1kkaa}
\end{figure}

The outer horizon covers the singularity inside the black hole. Hence, the location of the outer horizon plays an important role in the weak cosmic censorship conjecture. If the location of the outer horizon is irreducible, we ensure that the horizon cannot disappear, and the weak cosmic censorship conjecture is valid. Under Eq.\,(\ref{eq:DR2}), the initial state $(M,Q)$ changes to the final state $(M+dM,Q+dQ)$. In this context, an important issue should be mentioned. In gravity's rainbow, the energy $E$ of a particle causes modifications about the metric\cite{Magueijo:2002xx}. This is represented by the rainbow effect, and the necessary condition for the rainbow effect is existence of the particle seen by the asymptotic observer. Here, the charged particle lies on the black hole background before its absorption, so the energy of the particle induces the modification originated from gravity's rainbow. Then, we should consider the rainbow effect by the energy $E$ of the particle in the initial state. In the final state, the situation changes. The particle is now absorbed into the black hole and passes through the horizon, so the outside observer cannot detect the particle. This implies that the spacetime is the ordinary black hole without the rainbow effect to the observer, because there is no particle causing the rainbow effect. Instead, the mass and charge of the black hole change as much as those of the particle in the initial state. Depending on the change, the metric function of the final state becomes $H(r) = 1- 2(M+dM)/r + (Q+dQ)^2/r^2$. Thus, the initial and final outer horizons $r_{\rm H,inital}$ and $r_{\rm H,final}$ are solutions to
\begin{align}
F(r_{\rm H,inital})=0,\quad H(r_{\rm H,final})=0,
\end{align}
where $F(r)$ and $H(r)$ are functions of $g^{rr}$ with and without gravity's rainbow, respectively. When we assume that the final state is also a non-extremal black hole, the location of the horizon changes to
\begin{align}\label{eq:dFrH}
dr_{\rm H}&= r_{\rm H,final}-r_{\rm H,inital}\\
&=(M+dM)+\sqrt{(M+dM)^2-(Q+dQ)^2} - \left(M+\sqrt{M^2-g(dM) Q^2}\right) \nonumber\\
&= \frac{4M-\eta Q^2 + 4\sqrt{M^2-Q^2}}{4\sqrt{M^2-Q^2}}|p^r| 
-\frac{Q^3 \eta}{4\left(M^2-Q^2+M\sqrt{M^2-Q^2}\right)}dQ,\nonumber
\end{align}
which is obtained in the first order of variables. Interestingly, the change in the outer horizon depends on its charge $q$. Hence, if the particle has a sufficiently large charge, the outer horizon can decrease due to particle absorption. Further, when such decrease occurs in an extremal black hole, there is a possibility that the weak cosmic censorship conjecture can be invalid owing to overcharging. This aspect will be investigated in detail in the case of an extremal black hole. Here, it should be noted that the decrease in $r_\text{H}$ is due to the rainbow effect. In the limit of $\eta \rightarrow 0$ without the rainbow effect, the change in the outer horizon is only proportional to $|p^r|$ and always increases under particle absorption. The condition under which the particle decreases the outer horizon is obtained in terms of the inequality
\begin{align}\label{eq:inequality09}
\frac{q}{|p^r|} > \frac{-4Q^2+\left(M+\sqrt{M^2-Q^2}\right)\left(8M-\eta Q^2\right)}{\eta Q^3},
\end{align}   
where we assume that the charge $Q$ of black hole is positive. The behavior of $dr_\text{H}$ is shown in detail in Fig.\,\ref{fig:fig1kkaa}.
The negative regions for $dr_\text{H}$ increase for a large $\eta$, as shown in Fig.\,\ref{fig:fig1kkaa}. However, even if $\eta$ is small, the negative regions do not vanish, because the rainbow effect causes the decrease of the horizon. The outside of the black dashed lines are obtained according to Eq.\,(\ref{eq:inequality09}).

According to Eq.\,(\ref{eq:dFrH}), the horizon radius depends on the radial momentum and electric charge of the particle in the presence of the rainbow effect. However, the entropy is expected to be irreducible, because the second law of thermodynamics ensures that the entropy increases in an irreversible process, such as, particle absorption. Here, we investigate the change in the entropy with the rainbow effect. For the same reason stated in the previous analysis, the initial state of the black hole $(M,Q)$ is assumed to be the metric of Eq.\,\eqref{eq:metric_final}, including the rainbow effect. Then, the entropy of the initial black hole $S_\text{i}$ is given by Eq.\,(\ref{eq:TH}). After the black hole absorbs the particle, there is no particle in the spacetime, and hence the rainbow effect should also be removed in the final state. The final black hole is now of the form $(M+dM,Q+dQ)$ according to Eq.\,(\ref{eq:DR2}), and its metric follows Eq.\,(\ref{eq:E_metric}). Further, this changes the final entropy $S_\text{f}$ to Eq.\,(\ref{eq:T_RN}) without the rainbow effect. The change in the entropy is written as 
\begin{equation}\label{eq:S_f-S_i}
dS=S_{\rm f}(M+dM,Q+dQ,r_\text{H}+dr_\text{H}) - S_{\rm i}(M,Q,r_\text{H}).
\end{equation}
Then, the change in the entropy is obtained as
\begin{align}\label{eq:chentp07}
dS &=  \pi (r_{\rm H}+d r_{\rm H})^2 - \frac{\pi r_{\rm H}^2}{g(dM)^2} \\
&= \frac{\pi |p^r|}{2Q^2 -2M\left(M+\sqrt{M^2-Q^2}\right)} \left[ 8\eta M^4  + Q^2 \left(\eta Q^2+\left(4 - \eta Q \frac{dQ}{|p^r|}\right)\sqrt{M^2-Q^2}\right) \right.  \notag\\
& -4M^2 \left(2 \eta Q^2+\left(4 -\eta Q\frac{dQ}{|p^r|}\right)\sqrt{M^2-Q^2} \right)  
 +MQ^2\left(-3\eta Q \frac{dQ}{|p^r|} +12-4\eta\sqrt{M^2-Q^2}\right) \notag \\
&\left.  +4M^3 \left(\eta Q \frac{dQ}{|p^r|} -4+2\eta\sqrt{M^2-Q^2}\right)  \right].\nonumber
\end{align} 
Although particle absorption is an irreversible process, Eq.\,(\ref{eq:chentp07}) implies that the entropy can increase or decrease in accordance with the radial momentum and electric charge of the particle. Without the rainbow effect, $\eta\rightarrow 0$, the change in the entropy becomes irreducible.
\begin{figure}[h]
  \begin{center}
\subfigure[{~$Q-\frac{q}{|p^r|}$ diagram for $\eta = 0.02$}]{
  \includegraphics[width=0.321\textwidth]{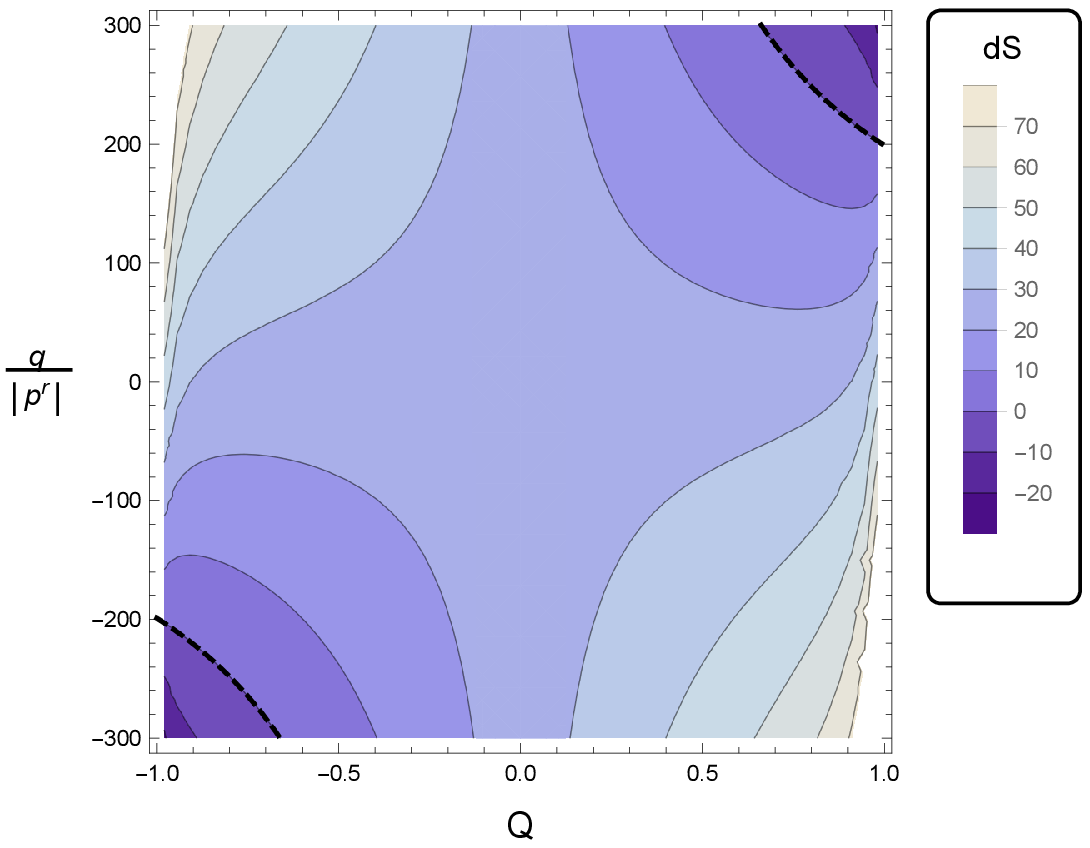}\label{fig:fig1a}}
\subfigure[{~$Q-\frac{q}{|p^r|}$ diagram for $\eta = 0.5$}]{
 \includegraphics[width=0.321\textwidth]{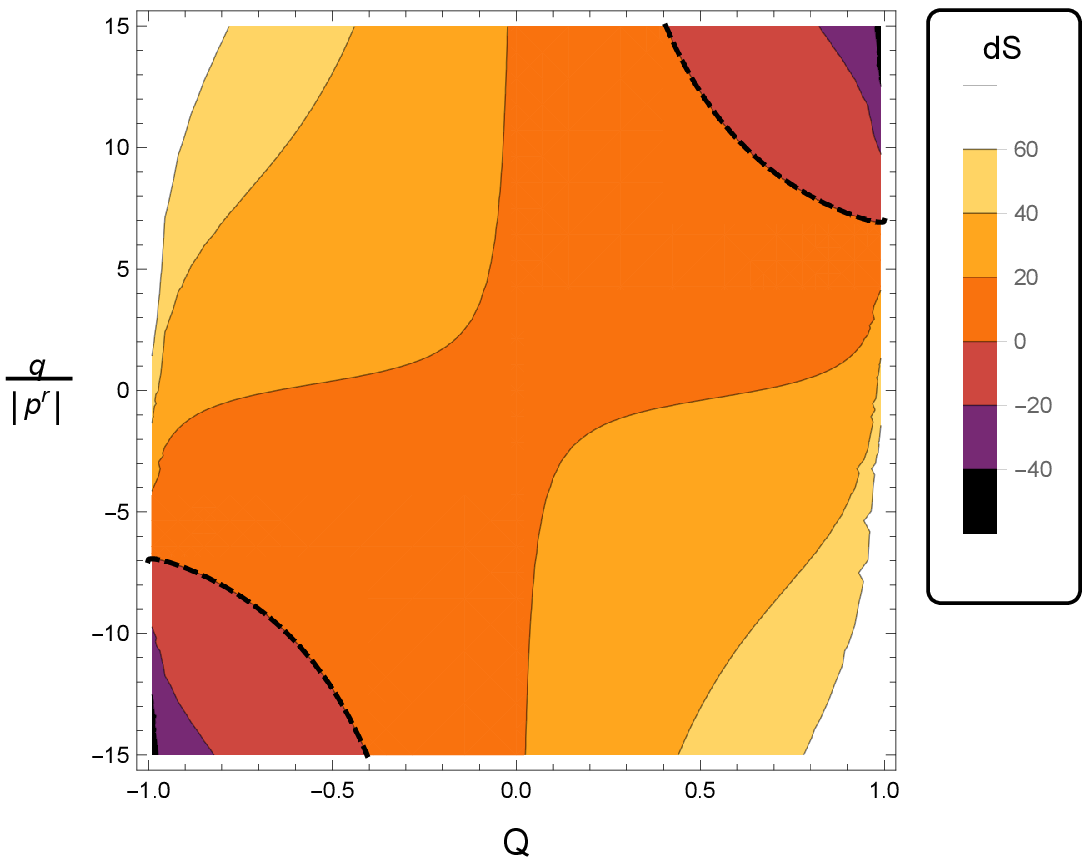}\label{fig:fig1b}}
\subfigure[{~$Q-\frac{q}{|p^r|}$ diagram for $\eta = 0.9$}]{
 \includegraphics[width=0.321\textwidth]{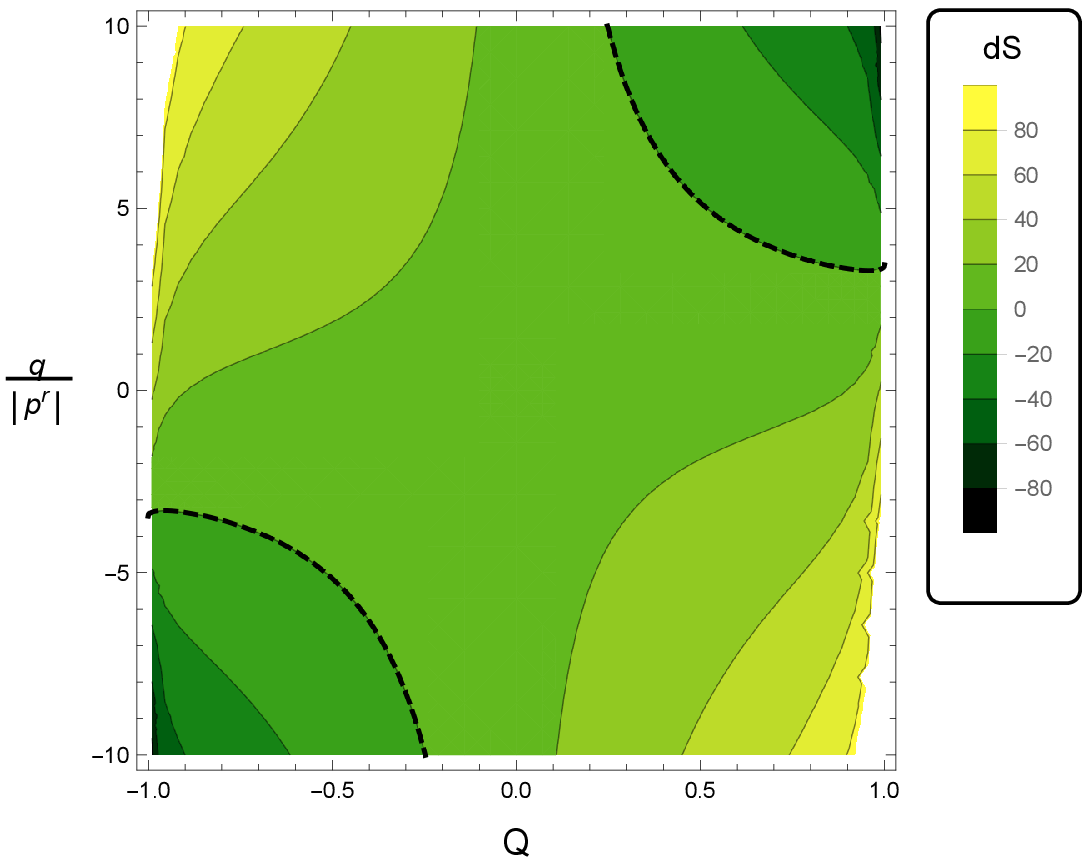}\label{fig:fig1c}}
  \end{center}
  \caption{Changes in the entropy $dS$ in $Q-\frac{q}{|p^r|}$ diagrams of $M=1$ for a given $\eta$.}\label{fig:fig1}
\end{figure}
Hence, the rainbow effect occurs with the violation of the second law of thermodynamics under charged particle absorption. Because the change in the entropy is divergent when the black hole approaches the extremal black hole, we should investigate detailed behaviors of the extremal black hole by a different method. Further, we can obtain the range of the radial momentum and electric charge of the particle to decrease the entropy of the system.
\begin{align}
\frac{q}{|p^r|}>\frac{4}{\eta Q}+\frac{ 8 M^4 - 8 M^2 Q^2+ Q^4 +\sqrt{M^2-Q^2} \left(8 M^3- 4MQ^2 \right)}{ Q\left(-4M^3+3MQ^2-(4M^2-Q^2)\sqrt{M^2-Q^2}\right)}. 
\end{align}
where $Q$ is assumed to positive. The detailed changes in the entropy are shown in Fig.\,\ref{fig:fig1}. In a small rainbow effect proportional to $\eta$ in Fig.\,\ref{fig:fig1a}, the negative regions of $dS$ correspond to relatively large values of $q$ and $Q$. Even if we consider an infinitesimally small rainbow effect, such as, $\eta\ll1$, the entropy of an extremal black hole has a possibility to decrease under particle absorption including a large electric charge. This implies that the second law of thermodynamics can be violated due to the rainbow effect. As the rainbow effect $\eta$ increases, the violation appears in boarder regions in Fig.\,\ref{fig:fig1b} and (c). In other words, the boundaries of negative changes in entropy, denoted using the black dashed lines in Fig.\,\ref{fig:fig1}, occupy broader areas for large values of $\eta$. Further, the divergence of $dS$ in Eq.\,(\ref{eq:chentp07}) is presented in the region of $|Q|\sim 1$ in Fig.\,\ref{fig:fig1}. Because the violation of the second law of thermodynamics implies the decrease of the horizon area, if the horizon of the extremal black hole becomes small, it can disappear owing to overcharging beyond the extremal condition. Thus, we need to investigate the extremal black hole in the context of the weak cosmic censorship conjecture.

\section{Violation of Weak Cosmic Censorship in Extremal Black Hole}\label{sec4}

The extremal black hole has the maximum charge for a given mass. Hence, by adding a particle, if the electric charge becomes greater than the mass of the black hole, the black hole is overcharged beyond the extremal condition. Further, the horizons will disappear, and the weak cosmic censorship conjecture will be invalid. To investigate the validity of the conjecture, we estimate the final state from the initial state of an extremal black hole under particle absorption.  However, we cannot use the same method applied in Eqs.\,(\ref{eq:dFrH}) and (\ref{eq:chentp07}), because the final state can be a naked singularity where Eqs.\,(\ref{eq:dFrH}) and (\ref{eq:chentp07}) are invalid. To estimate the final state, we have to focus on the analytical structure of the metric function $F(r)$, which is well defined in the cases of both black hole and naked singularity.

\begin{figure}[h]
  \begin{center}
\subfigure[{The initial extremal black hole of $(M,Q)$.}]{
  \includegraphics[width=0.4\textwidth]{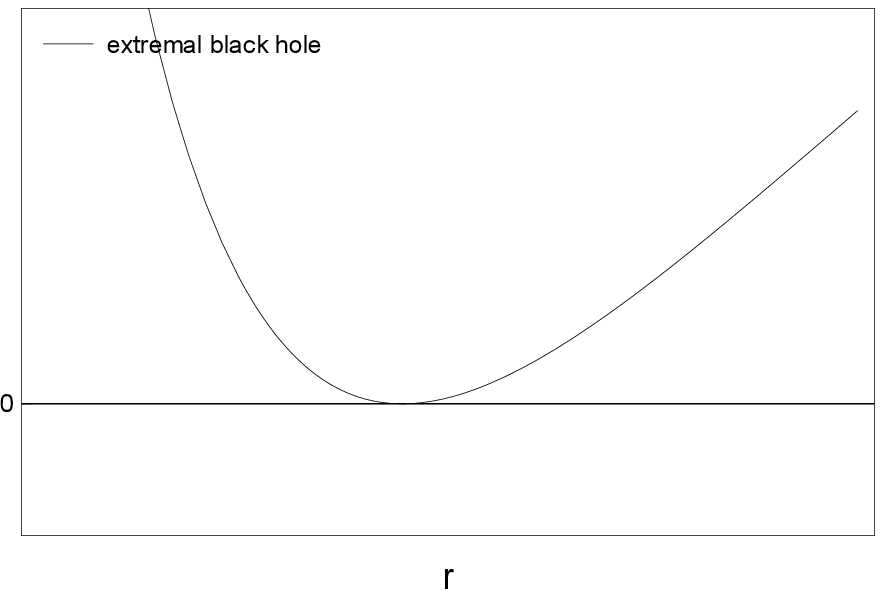}\label{fig03a} } \quad
\subfigure[{The possible final states of $(M+dM,Q+dQ)$.}]{
 \includegraphics[width=0.4\textwidth]{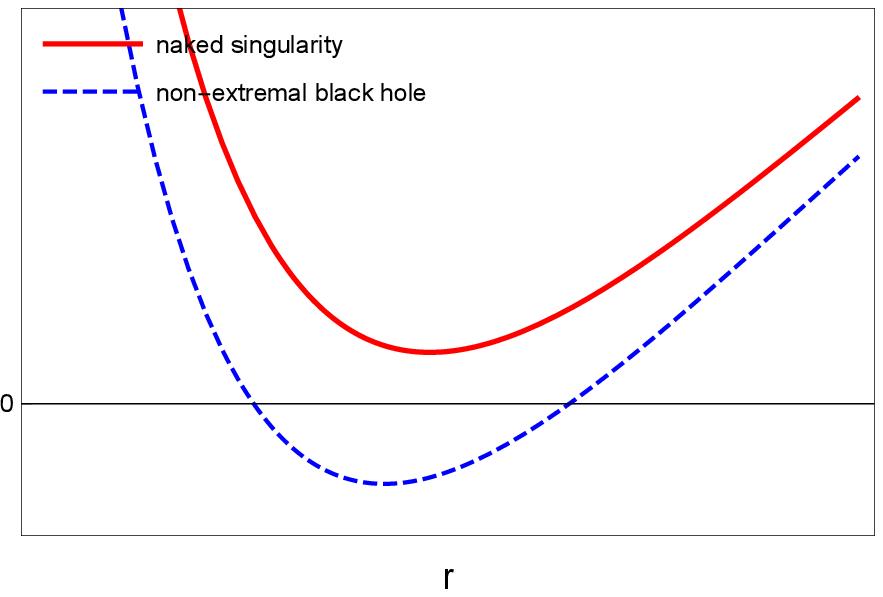}\label{fig03b}}
  \end{center}
  \caption{The function $F(r)$ in initial and final states.}\label{fig03}
\end{figure}

The function $F(r)$ of the extremal black hole $(M,Q)$ has only one minimum point located at the horizon $r_{\rm e}$, which satisfies
\begin{align}\label{eq:}
F(M,Q,r)|_{r=r_{\rm e}}=0,\quad \partial_r F(M,Q,r)|_{r=r_{\rm e}}=0, \quad (\partial_r)^2 F(M,Q,r)|_{r=r_{\rm e}} > 0.
\end{align}
This implies that the inner and outer horizons are coincident and located at the minimum point, as shown in Fig.\,\ref{fig03a}.
Further, the minimum value of $F(r)$ is zero in the initial state. Under charged particle absorption, the function $F(r)$ of the initial state becomes $H(r)$ of the final state in $(M+dM,Q+dQ)$, because there is no particle providing the rainbow effect in the final state.  Hence, the minimum point and value are also moved in the final state. From the moved minimum value, we can estimate the final state, as shown in Fig.\,\ref{fig03b}. When the minimum value is positive in the final state, the function $F(r)$ has no solution corresponding to the inner or outer horizon, as indicated by the red line in Fig.\,\ref{fig03b}. Then, the singularity is not covered by a horizon, and the conjecture is invalid. If the final state is still a black hole, the function $F(r)$ has a negative minimum value and horizons as shown by the blue line in Fig.\,\ref{fig03b}. Thus, we can estimate the final state from the sign of the minimum value. The change in the minimum value is
\begin{align}\label{eq:changeminimum}
dF_\text{min}&=H(M+dM,Q+dQ,r_\text{e}+dr_\text{e})-F(M,Q,r_\text{e})\\
&=-\frac{2|p^r|}{r_{\rm e}}+\frac{\eta}{2}(dQ+|p^r|),\nonumber
\end{align}
where we consider up to the first order of expansion. Without the rainbow effect $\eta= 0$ in Eq.\,(\ref{eq:changeminimum}), the minimum value has a negative sign, which implies non-extremal black hole, as shown by the blue dashed line in Fig.\,\ref{fig03b}. 
\begin{figure}[h]
  \begin{center}
\subfigure[{~$r_{\rm e}-\frac{q}{|p^r|}$ diagram for $\eta = 0.02$}]{
  \includegraphics[width=0.32\textwidth]{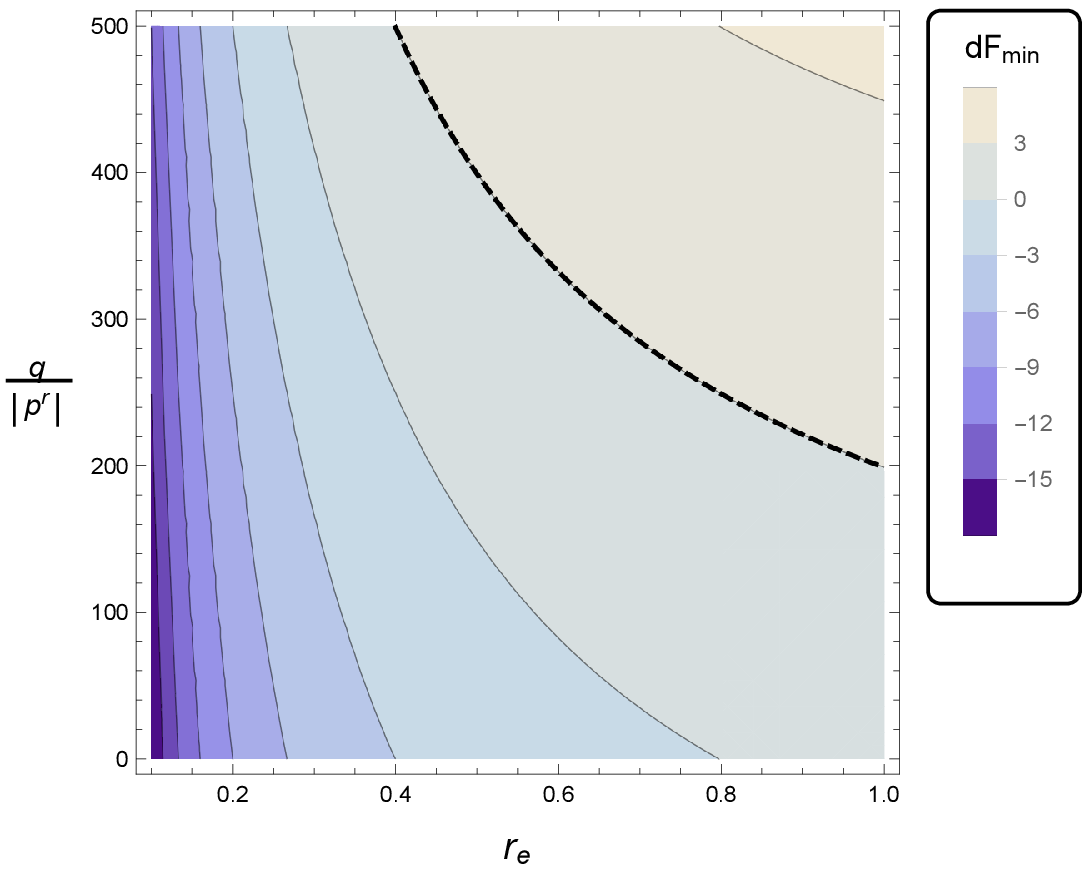}\label{fig:WCCa}}
\subfigure[{~$r_{\rm e}-\frac{q}{|p^r|}$ diagram for $\eta = 0.5$}]{
 \includegraphics[width=0.32\textwidth]{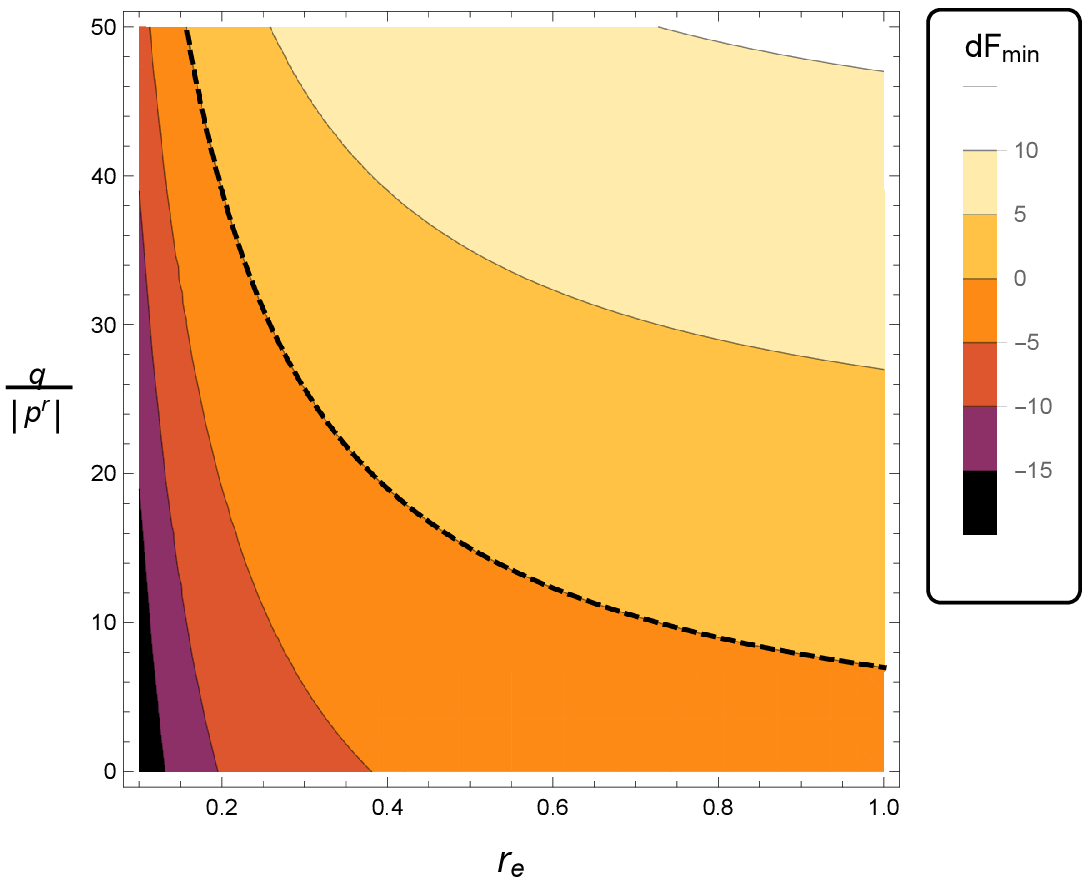}\label{fig:WCCb}}
\subfigure[{~$r_{\rm e}-\frac{q}{|p^r|}$ diagram for $\eta = 0.9$}]{
 \includegraphics[width=0.32\textwidth]{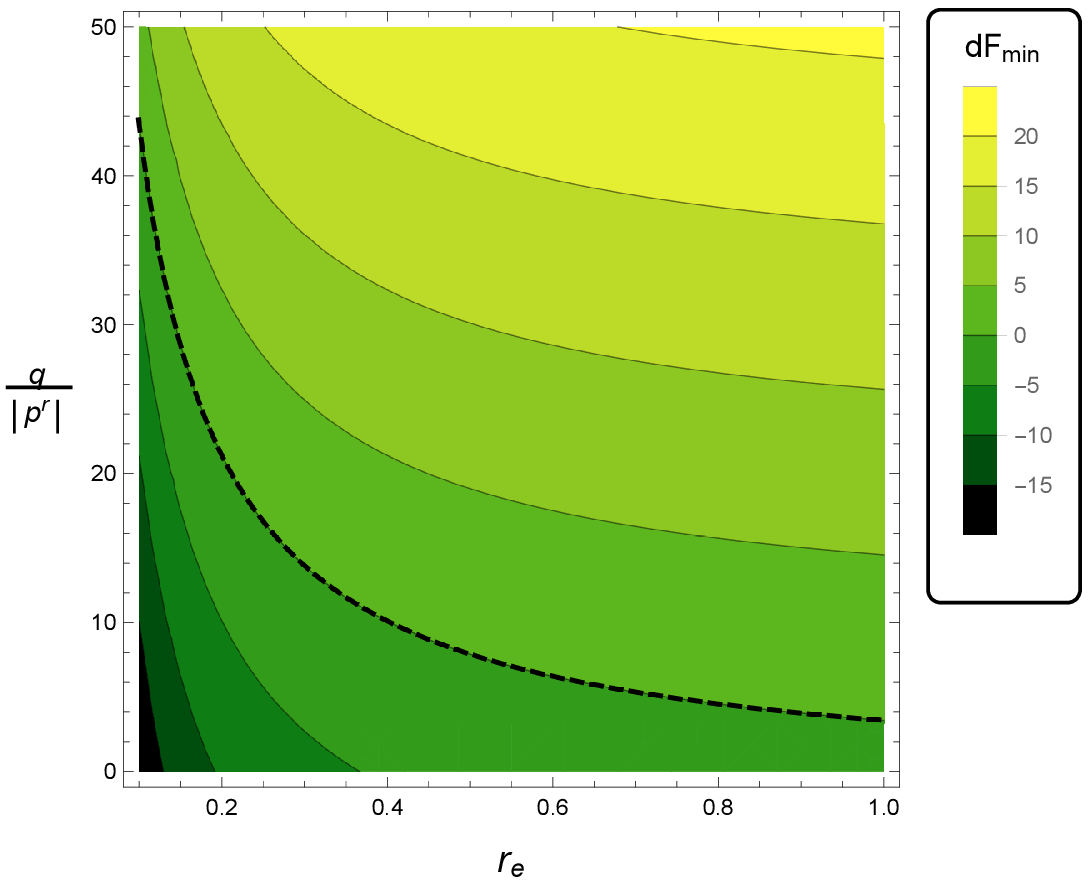}\label{fig:WCCc}}
  \end{center}
  \caption{Changes in the minimum value of the function $dF_\text{min}$ in $r_\text{e}-\frac{q}{|p^r|}$ diagrams for a given $\eta$.} \label{fig:WCC}
\end{figure}
This is already reported in previous literature\cite{Wald:1974ge,Isoyama:2011ea,Sorce:2017dst}, and the weak cosmic censorship is valid. However, with the rainbow effect, $\eta\not= 0$, interestingly, the sign of the minimum value depends on the radial momentum and electric charge of the particle. The change of the minimum value becomes positive, when
\begin{align}\label{eq:nakedsing}
\frac{q}{|p^r|}>\frac{4}{\eta r_\text{e}}-1,
\end{align}
where we assume $\eta \ll 1$. This implies that the highly charged particle with respect to its kinetic energy causes overcharging of the extremal black hole. Thus, the weak cosmic censorship conjecture is {\it invalid}. According to Eq.\,(\ref{eq:changeminimum}), the change in the minimum value $dF_\text{min}$ is shown in detail in Fig.\,\ref{fig:WCC} with respect to $\frac{q}{|p^r|}$ of the particle. As we expected, overcharging beyond the extremal condition becomes broader for a larger value of $\eta$ in comparison with Figs.\,\ref{fig:WCC}\,(a), (b), and (c). The largely charged particle tends to overcharge the extremal black hole. However, for a massive extremal black hole, overcharging can be caused by a particle with relatively small charge. The decrease of the black dashed line representing Eq.\,(\ref{eq:nakedsing}) clearly shows this behavior in Fig.\,\ref{fig:WCC}. This is based on the dependence on $r_\text{e}$ and $\eta$ in Eq.\,(\ref{eq:changeminimum}). The negative contribution of $|p^r|$ becomes small in a massive extremal black hole, but its positive contribution is substantial under the rainbow effect. As a result, the massive extremal black hole becomes a naked singularity for a relatively small charge of the particle for a large value of $\eta$. Therefore, the rainbow effect plays an important role in overcharging an extremal black hole.

\section{Summary}\label{sec5}

We have investigated violations in the second law of thermodynamics and weak cosmic censorship conjecture in an electrically charged black hole with gravity's rainbow. Among the sets of rainbow functions representing various aspects of MDRs, we choose one that is well consistent with the quantum-spacetime-phenomenology perspective\cite{AmelinoCamelia:1996pj, AmelinoCamelia:1997gz}, such as, the loop-quantum-gravity approach \cite{Gambini:1998it, Alfaro:2001rb, Sahlmann:2002qk, Smolin:2002sz, Smolin:2005cz}. Considering the rainbow functions, the charged black hole is modified to impose the rainbow effect from the MDR on its metric. Then, we have studied infinitesimal variations of the rainbow charged black hole caused by a charged particle to understand the effect of the MDR. The particle including the rainbow effect is assumed to change the mass and charge of the black hole as much as its own energy and charge, when it passes through the outer horizon. Here, we note a remarkable point about charged particle absorption. Because the rainbow effect is presented by the charged particle, the initial state is assumed to be the rainbow black hole, but the final state does not include the rainbow effect in its spacetime, because the particle does not exist due to absorption into the black hole. This concept differs from previous studies without gravity's rainbow. However, changes in the outer horizon and Bekenstein-Hawking entropy depend on the radial momentum and charge of the particle with the rainbow effect and do not occur without the rainbow effect. Hence, owing to the rainbow effect, the second law of thermodynamics is violated by the absorption of a largely charged particle. Further, because the change diverges when the initial state is assumed to be an extremal black hole, we investigated the conjecture for the case of extremal black hole. In consideration of the minimum value of the function $F(r)$, we have proven that the extremal black hole can be a naked singularity for a largely charged particle owing to the unstable horizon. Here, the rainbow effect plays an important role to establish this invalidity of the weak cosmic censorship conjecture. It is worth noting that violations of the second law of thermodynamics and weak cosmic censorship conjecture in the presence of gravity's rainbow are demonstrated for the first time in this study.

\vspace{10pt}

\noindent {\bf Acknowledgments}

\noindent 
This work was supported by the National Research Foundation of Korea (NRF) grant funded by the Korea government (MSIT) (NRF-2018R1C1B6004349). B.G. appreciates APCTP for its hospitality during completion of this work.

\end{document}